\documentclass[amsmath,amssymb,aps,prb,floatfix,reprint,longbibliography]{revtex4-2}
\usepackage{amsmath}
\usepackage{epsfig}
\usepackage{graphicx}
\usepackage{graphicx, color, epstopdf}
\usepackage{bm}
\usepackage{amssymb}
\usepackage{hyperref}
\usepackage{xcolor}
\usepackage{subfigure}
\usepackage{makecell}
\usepackage{tikz}
\usepackage{graphicx}   
\usepackage{amsfonts}   
\usepackage{bm}         
\usepackage{hyperref}   
\usepackage{xcolor}
\usepackage{cleveref}
\usepackage{appendix}
\usepackage{afterpage}

\hypersetup{hidelinks,
	colorlinks=true,
	allcolors=black,
	pdfstartview=Fit,
	breaklinks=true}
	
\definecolor{lime}{HTML}{A6CE39}
\DeclareRobustCommand{\orcidicon}{
\begin{tikzpicture}
\draw[lime, fill=lime] (0,0)
circle[radius=0.16]
node[white]{{\fontfamily{qag}\selectfont \tiny \.{I}D}};
\end{tikzpicture}
\hspace{-2mm}
}
\foreach \x in {A, ..., Z}{%
\expandafter\xdef\csname orcid\x\endcsname{\noexpand\href{https://orcid.org/\csname orcidauthor\x\endcsname}{\noexpand\orcidicon}}
}

\newcommand{\imi}{\mathrm{i}}

\begin{document}
\title{Anomalous exchange correlation of quasiparticles with entangled Nambu spinors}
\author{Hai-Dong Li}
\affiliation{National Laboratory of Solid State Microstructures, School of Physics,
  Jiangsu Physical Science Research Center, and Collaborative Innovation Center of Advanced Microstructures,
  Nanjing University, Nanjing 210093, China}
\author{Wei Chen \hspace{-1.5mm}\orcidA{}}
\email{Corresponding author: pchenweis@gmail.com}
\affiliation{National Laboratory of Solid State Microstructures, School of Physics,
  Jiangsu Physical Science Research Center, and Collaborative Innovation Center of Advanced Microstructures,
  Nanjing University, Nanjing 210093, China}
\date{\today}
\begin{abstract}
  Entanglement of spin degree of freedom can drastically alter the orbital exchange symmetry of electrons, switching their bunching and antibunching behaviors and the resultant current correlations in the Hanbury-Brown-Twiss interferometry.
  Here, we investigate the exchange correlation of quasiparticles with entanglement encoded in the Nambu spinors, or the electron-hole degree of freedom. In contrast to the conventional correspondence between spin entanglement and current correlation,  we find that singlet (triplet) entanglement of Nambu spinors results in suppressed (enhanced) current correlation. This effect arises because the charge degree of freedom itself encodes the entanglement. We propose implementing this phenomenon in the edge states of a quantum Hall system, where the entangled states of the Nambu spinors can be continuously tuned by gate voltages. Our study reveals a novel relationship between entanglement and charge correlations, offering an effective approach for detecting entanglement of Nambu spinors.
\end{abstract}

\maketitle


The exchange symmetry of identical particles can be effectively revealed through shot noise measurements in Hanbury-Brown-Twiss interferometry~\cite{eberlyCoherenceQuantumOptics2012, oliverHanburyBrownTwissType1999c}. Typically, Fermi statistics lead to antibunching behavior of electrons, thereby suppressing shot noise~\cite{buttikerScatteringTheoryThermal1990, buttikerScatteringTheoryCurrent1992, martinWavepacketApproachNoise1992a, oliverHanburyBrownTwissType1999b, hennyFermionicHanburyBrown1999a, liuQuantumInterferenceElectron1998}, while Bose statistics result in photon bunching and enhance shot noise beyond the classical limit~\cite{oliverHanburyBrownTwissType1999b}. The underlying principle is that the two-particle wavefunction, or more precisely, its spatial part, is symmetric for bosons and antisymmetric for fermions. Interestingly, when the wavefunction contains a nontrivial internal degree of freedom, such as spin or pseudo-spin, the exchange symmetry of its spatial part can be modified or even inverted, while the overall wavefunction retains its original exchange symmetry~\cite{sakuraiModernQuantumMechanics2020}. For example, in the case of an electron pair in a spin singlet state, the spatial part of the wavefunction becomes exchange symmetric, leading to a bunching behavior similar to that of two photons~\cite{burkardNoiseEntangledElectrons2000}. This property, where spin entanglement modifies the spatial correlations of identical particles, provides an effective method for probing electronic entanglement in solid-state physics~\cite{eguesShotNoiseSpinorbit2005b, eguesRashbaSpinOrbitInteraction2002, burkardNoiseEntangledElectrons2000, mazzaSpinFilteringEntanglement2013, engelElectronSpinsQuantum2001a, maitreEntanglement2DEGSystems2000}. Specifically, spin singlet and triplet entanglements correspond to electron bunching and antibunching, respectively, resulting in enhanced and suppressed shot noise~\cite{burkardNoiseEntangledElectrons2000}.

In this work, we uncover a novel scenario in which the correspondence between the spatial exchange correlation of quasiparticles and shot noise is reversed. Specifically, the bunching and antibunching of entangled quasiparticles lead to suppressed and enhanced shot noise, respectively, in contrast to the behavior observed for spin-entangled electrons. This is achieved by encoding entanglement in the electron-hole degree of freedom, or the Nambu spinors~\cite{luoEntanglementNambuSpinors2022}. Consequently, the exchange correlations in quasiparticle scattering and charge transport exhibit opposite behaviors.
\begin{figure*}
  \centering
  \includegraphics[width=0.75\textwidth]{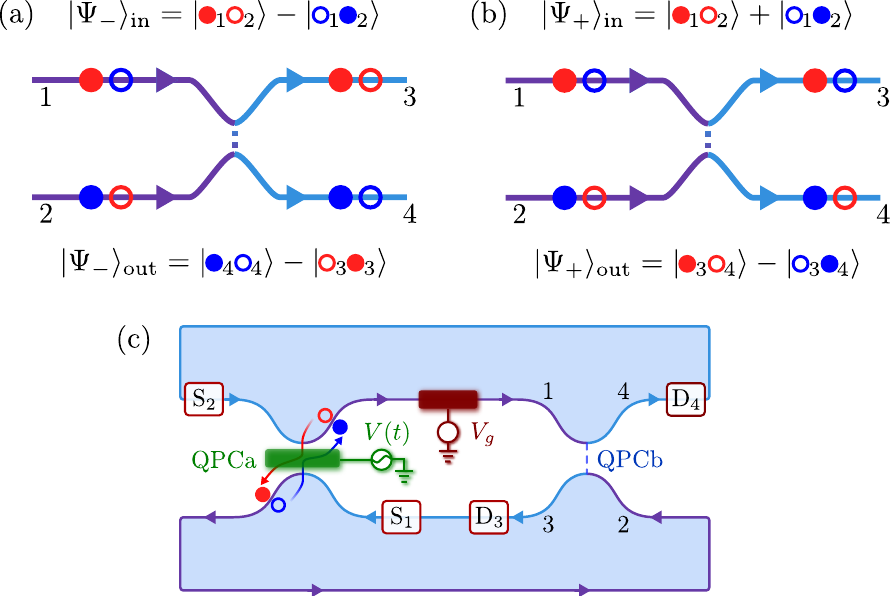}
  \caption{
    Scattering of quasiparticles at an ideal beam splitter with (a) singlet ($|\Psi_- \rangle_{\text{in}}$) and (b) triplet ($|\Psi_+ \rangle_{\text{in}}$) entanglement in the electron-hole degree of freedom. Quasiparticles are injected from channels 1 and 2 and exit through channels 3 and 4. Solid and hollow circles represent electrons and holes, respectively, while blue and red colors indicate quantum superposition of the states. A singlet injected state results in bunching behavior, where electrons and holes preferentially exit through the same channel. In contrast, a triplet injected state exhibits antibunching behavior, where electrons and holes preferentially exit through different channels. We denote the outgoing states after scattering as $|\Psi_- \rangle_{\text{out}}$ and $|\Psi_+ \rangle_{\text{out}}$ for the singlet and triplet cases, respectively.
    (c) The proposed setup constructed on the chiral edge states (indicated by arrowed lines) in the integer quantum Hall regime. Similar to (a) and (b), solid and hollow circles in the diagram represent electrons and holes, with distinct colors denoting superposed states. Entangled pairs are generated by the periodic potential \(V(t)\) within the point contact region denoted by QPCa. Both sources \((\mathrm{S}_1, \mathrm{S}_2)\) and drains \((\mathrm{D}_{3}, \mathrm{D}_{4})\) are grounded. A voltage gate \(V_{g}\) covers one of the output channels, modulating the entangled state. The point contact QPCb acts as a beam splitter with adjustable transparency, and the numbering there corresponds to that in (a) and (b).
  }
  \label[figure]{fig.1}
\end{figure*}
To be specific, a Bogoliubov quasiparticle is denoted as \( |\psi_{\mathrm{B}}\rangle = \left( \alpha b_\mathrm{e}^\dag + \beta b_\mathrm{h}^\dag \right) |0\rangle \), which is a superposition of the electron and hole components with amplitudes \( \alpha \) and \( \beta \), respectively, expressed by the creation operators \( b_{\mathrm{e},\mathrm{h}}^\dag \) acting on the vacuum state \( |0\rangle \). Mathematically, this state resembles an electronic spin state \( |\psi_{\mathrm{S}}\rangle = \left( \alpha b_\uparrow^\dag + \beta b_\downarrow^\dag \right) |0\rangle \), with the replacement \( \uparrow (\downarrow) \to e (h) \) for the spinors.
Now, consider a pair of Bogoliubov quasiparticles whose electron and hole components are maximally entangled~\cite{luoEntanglementNambuSpinors2022}, as shown in Fig.~\ref{fig.1}. This entangled state can be expressed as
\[
  |\Psi_{\pm}\rangle_{\text{in}} = \frac{1}{\sqrt{2}}\left( b_{1\mathrm{e}}^\dag b_{2\mathrm{h}}^\dag
  \pm b_{1\mathrm{h}}^\dag b_{2\mathrm{e}}^\dag \right) |0\rangle,
\]
where \( b_{i \alpha}^\dag \) is the creation operator of the \( \alpha~ (= \mathrm{e}, \mathrm{h})\)-type quasiparticle in the \( i \) (= 1, 2) channel, and ``\( \pm \)'' correspond to the triplet and singlet states of the pseudo-spin, respectively. This means that if an electron is created in channel 1, a hole is simultaneously created in channel 2, and vice versa. The physical interpretation of these entangled states becomes clearer when comparing them to their spin counterparts
\(
|\Psi_{\mathrm{S}}^{\pm}\rangle = \frac{1}{\sqrt{2}} \left( b_{1\uparrow}^\dag b_{2\downarrow}^\dag
\pm b_{1\downarrow}^\dag b_{2\uparrow}^\dag \right) |0\rangle.
\)

To illustrate the anomalous correspondence between the bunching/antibunching effects and the shot noise induced by this novel type of entanglement, consider two entangled quasiparticles incident from channels 1 and 2 of the Hanbury-Brown-Twiss interferometer, which are then scattered to the outgoing channels 3 and 4. For an ideal beam splitter, an incident particle has an equal probability of transmitting into channels 3 and 4. A straightforward calculation shows that for the incident singlet state \( |\Psi_- \rangle_{\text{in}} \), the corresponding outgoing state is given by
\[
  |\Psi_- \rangle_{\text{out}} = \frac{1}{\sqrt{2}} \left( b_{4\mathrm{e}}^\dag b_{4\mathrm{h}}^\dag
  - b_{3\mathrm{h}}^\dag b_{3\mathrm{e}}^\dag \right) |0\rangle.
\]
This indicates that the outgoing quasiparticles are always bound together to enter either channel 3 or 4, displaying the typical bunching effect; see Fig.~\ref{fig.1}(a). However, the correlated quasiparticles always have opposite charges, leading to zero current and shot noise in channels 3 and 4. This is in stark contrast to the results for spin-entangled electrons, where the strong bunching effect gives rise to the strongest shot noise~\cite{burkardNoiseEntangledElectrons2000}.
Conversely, for a triplet incident state \( |\Psi_+ \rangle_{\text{in}} \) passing through the beam splitter, the outgoing state is
\[
  |\Psi_+ \rangle_{\text{out}} = \frac{1}{\sqrt{2}} \left( b_{3\mathrm{e}}^\dag b_{4\mathrm{h}}^\dag
  - b_{3\mathrm{h}}^\dag b_{4\mathrm{e}}^\dag \right) |0\rangle,
\]
meaning that the two quasiparticles always exit through different output channels; see Fig.~\ref{fig.1}(b). Remarkably, this antibunching behavior generates the strongest shot noise, once again due to the opposite charges carried by the correlated electron and hole components. Such anomalous current correlations provide an effective method for probing the entanglement of Nambu spinors~\cite{luoEntanglementNambuSpinors2022}.
We propose implementing these novel effects in the Hanbury-Brown-Twiss interferometer constructed by quantum point contacts (QPCs) based on quantum Hall edge states~\cite{weiszElectronicQuantumEraser2014a, nederInterferenceTwoIndistinguishable2007b, jiElectronicMachZehnder2003b, hennyFermionicHanburyBrown1999b}, as shown in Fig.~\ref{fig.1}(c). The entangled Nambu spinor can be generated using a driving gate~\cite{bisogninQuantumTomographyElectrical2019a, duboisMinimalexcitationStatesElectron2013c, vanevicElectronElectronholeQuasiparticle2016a, gaborExtremelyEfficientMultiple2009a, feveDemandCoherentSingleElectron2007}, and entangled states can be conveniently tuned via gate voltages.

The proposed Hanbury-Brown-Twiss interferometer is illustrated in Fig.~\ref{fig.1}(c), constructed from the chiral edge states of the integer quantum Hall system.
The many-body state describing the electrons incident from \( \mathrm{S}_1 \) and \( \mathrm{S}_2 \) is expressed as
\begin{equation}\label{eq.comingstate}
  |\Psi\rangle_{\mathrm{in}} = \prod_{E < 0} a^{\dagger}_{1 \mathrm{e}}(E) a^{\dagger}_{2 \mathrm{e}}(E)|0\rangle,
\end{equation}
where \(a^{\dagger}_{1 \mathrm{e},2 \mathrm{e}}(E)\) denotes the creation operators for incident electrons with energy \(E\), measured relative to the Fermi level (set to zero). Here, \(|0\rangle\) represents the electron vacuum state.
Entangled quasiparticle pairs are excited in the QPCa region by a periodic driving potential \(V(t) = V_{0} + V_{1} \cos(\omega t)\), and propagate along the chiral edge states, as indicated by the arrows. This induces a periodic variation in the transmission amplitude of the scattering matrix, allowing particles passing through the QPCa to absorb or emit multiple energy quanta, \(n \hbar \omega\) \cite{samuelssonTwoParticleAharonovBohmEffect2004}. Within the framework of Floquet scattering theory \cite{moskaletsDissipationNoiseAdiabatic2002b, moskaletsScatteringMatrixApproach2011a}, this process can be described by the energy-dependent scattering m atrix \(S_{\mathrm{a}}(E, E_n)\) as
\begin{equation}\label{eq.scat.matrix}
  \left(\begin{array}{l}
      b_{1}(E) \\
      b_{2}(E)
    \end{array}\right)=\sum_{n} S_{\mathrm{a}}\left(E, E_n\right)\left(\begin{array}{l}
      a_1\left(E_n\right) \\
      a_2\left(E_n\right)
    \end{array}\right),
\end{equation}
where \(a_{i}(E_n)\) and \(b_{i}(E)\) represent the incoming and outgoing states, respectively, with $E_n=E+n\hbar \omega$. A gate voltage \(V_g\) is applied to one of the paths originating from QPCa, where quasiparticles undergo phase modulations. Because electrons and holes carry opposite charges, they experience opposite phase shifts under the influence of \(V_g\), allowing for electrical manipulation of the entangled states.
QPCb servers as a quasiparticle collider, where the entangled quasiparticles collide and undergo a Hanbury-Brown-Twiss-type interference. The outgoing states are then directed into two drains \(\mathrm{D}_{3}\) and \(\mathrm{D}_{4}\).

Here, we focus on the limiting case of a low driving potential and a low driving frequency. In this regime, Eq.~(\ref{eq.scat.matrix}) can be simplified. In particular, when \(V_1\) is small, it is sufficient to consider the static scattering and single-photon-assisted scattering, given by
\begin{equation*}
  \begin{aligned}
     & S_{\mathrm{a}}\left(E, E\right) \equiv
    s_0 = \left[r, t'; t, r'\right],                  \\
     & S_{\mathrm{a}}\left(E, E_{\pm 1}\right) \equiv
    s_{\pm} = \left[\delta r_{\pm}, \delta t'_{\pm}; \delta t_{\pm}, \delta r'_{\pm}\right].
  \end{aligned}
\end{equation*}
Furthermore, a low driving frequency allows us to consider the adiabatic regime, \(\omega \delta \tau \ll 1\), with \(\delta \tau\) the time that a quasiparticle spends in the QPCa region. In this case, \(s_{\pm}\) can be expressed in terms of \(s_0\) as \(s_{\pm} \equiv s_1 = V_1 \left(\frac{\partial s_0}{\partial V_0}\right) / 2 = \left[\delta r, \delta t'; \delta t, \delta r'\right]\) \cite{moskaletsDissipationNoiseAdiabatic2002b}.

Ultimately, the incoming state is scattered into a superposition of outgoing states with different energies as
\begin{equation}\label{eq.trans}
  \left(a_1^{\dagger}(E), a_2^{\dagger}(E)\right)=
  \sum_{n=0,\pm 1}
  \left(b_1^{\dagger}(E_{n}), b_2^{\dagger}(E_{n})\right)
  S_{\mathrm{a}}\left(E, E_n\right).
\end{equation}
By substituting Eq.~(\ref{eq.trans}) into the incoming state Eq.~(\ref{eq.comingstate}), the outgoing state can be obtained as
\begin{equation}\label{state}
  \begin{aligned}
     & \left|\Psi_{\mathrm{out}}\right\rangle
    = |\bar{0}\rangle + |\mathrm{E}\rangle + |\mathrm{N}\rangle,                                   \\
     & |\mathrm{E}\rangle
    = \int_{-\hbar \omega}^0 d E                                                                   \\
     & \times \left( g_{12} b_{1\mathrm{e}}^{\dagger}(E_1) b_{2\mathrm{h}}^{\dagger}(-E)
    - g_{21} b_{1\mathrm{h}}^{\dagger}(-E) b_{2\mathrm{e}}^{\dagger}(E_1) \right) |\bar{0}\rangle, \\
     & | \mathrm{N}\rangle
    = \int_{-\hbar \omega}^0 d E                                                                   \\
     & \times \left( g_{11} b_{1\mathrm{e}}^{\dagger}(E_1) b_{1\mathrm{h}}^{\dagger}(-E)
    + g_{22} b_{2\mathrm{e}}^{\dagger}(E_1) b_{2\mathrm{h}}^{\dagger}(-E) \right) |\bar{0}\rangle.
  \end{aligned}
\end{equation}
where \( |\bar{0}\rangle \equiv \prod_{E<0} b_{1}^{\dagger}(E) b_{2}^{\dagger}(E) |0\rangle \) represents the vacuum state for the outgoing quasiparticles at QPCa, $|\mathrm{E}\rangle$ is the nonlocal entangled state, and $|\mathrm{N}\rangle$ corresponds to a locally excited electron-hole pairs without nonlocal entanglement that is denoted as the normal state.
Due to Pauli exclusion, the $s_-$ terms do not contribute to the outgoing states.
The particle-hole transformation is employed as follows
\begin{equation*}
  \begin{aligned}
    b_{i\mathrm{e}}(E) \equiv b_{i}(E),
    b_{i\mathrm{h}}(E) \equiv b_{i}^{\dagger}(-E).
  \end{aligned}
\end{equation*}
To simplify the notation, the coefficients \(\{g_{ij}\}\) in Eq.~\eqref{state} have been expressed as the elements of the matrix defined as
\begin{equation}\label{eq.gMatrix}
  g \equiv s^{\dagger}_0 s_{1} =
  \begin{pmatrix}
    \delta r r^* + \delta t' t^{\prime *} & \delta r t^* + \delta t' r^{\prime *} \\
    \delta t r^* + \delta r' t^{\prime *} & \delta t t^* + \delta r' r^{\prime *}
  \end{pmatrix}.
\end{equation}
The unitarity of the scattering matrix imposes constraints on the $g$ matrix, specifically \(g = - g^{\dagger}\), with additional constraints provided in Appendix~\ref{sec.appendix1}. Taking these constraints into account, the $g$ matrix in Eq.~\eqref{eq.gMatrix} can be expressed in the following form
\begin{equation}\label{eq.gMatrixFinal}
  g =
  \begin{pmatrix}
    \mathrm{i} \rho_{\mathrm{N}}                        & \rho_{\mathrm{E}} \mathrm{e}^{\mathrm{i} \alpha} \\
    - \rho_{\mathrm{E}} \mathrm{e}^{-\mathrm{i} \alpha} & - \mathrm{i} \rho_{\mathrm{N}}
  \end{pmatrix},
\end{equation}
where \(\rho_{\mathrm{N},\mathrm{E}}\) and \(\alpha\) are all real numbers and the subscripts $\mathrm{N,E}$ indicate the amplitudes of
the normal and entangled states, respectively.

The phase modulation to the electron and hole components by the gate voltage \(V_g\) can be described by
\begin{equation*}
  b_{\mathrm{e}} \rightarrow b_{\mathrm{e}} \mathrm{e}^{\mathrm{i} \phi}, \quad b_{\mathrm{h}} \rightarrow b_{\mathrm{h}} \mathrm{e}^{-\mathrm{i} \phi},
\end{equation*}
where \(\phi = e V_g L_g /(\hbar v)\), with \(L_g\) the length of the gating region and \(v\) the velocity of the quasiparticle. The opposite phase modulation stems from the opposite charges carried by the electron and hole.
Importantly, the gate voltage can effectively modulate the \( |\mathrm{E}\rangle \) state because the entanglement takes place in the charge degree of freedom. In contrast, the \( |\mathrm{N}\rangle \) state is unaffected by the gate voltage, as the phases accumulated during the propagation of electron and hole along the same edge channel cancel each other out. The final form of the entangled state after the phase modulation by the voltage gate is then
\begin{equation}\label[equation]{eq.stateEntangle}
  \begin{aligned}
    |\mathrm{E}\rangle
     & = \rho_{\mathrm{E}} \int_{-\hbar \omega}^0 d E                                                                  \\
     & \times \left( \mathrm{e}^{\mathrm{i} \tilde{\phi}} b_{1\mathrm{e}}^{\dagger}(E_1) b_{2\mathrm{h}}^{\dagger}(-E)
    + \mathrm{e}^{-\mathrm{i} \tilde{\phi}} b_{1\mathrm{h}}^{\dagger}(-E) b_{2\mathrm{e}}^{\dagger}(E_1) \right) |\bar{0}\rangle,
  \end{aligned}
\end{equation}
where \(\tilde{\phi} = \alpha - \phi\). Notably, \(\tilde{\phi} = 0\) results in a triplet state, while \(\tilde{\phi} = \pi / 2\) leads to a singlet state.
It serves as the incident state for the Hanbury-Brown-Twiss interferometer at QPCb in Fig.~\ref{fig.1}(c).

The nonlocal entangled Nambu spinors can be revealed by the current-current correlation, or shot noise. The current operator in the detection regions is defined as~\cite{buttikerScatteringTheoryCurrent1992, buttikerScatteringTheoryThermal1990}:
\begin{equation}\label{eq.current}
  \begin{aligned}
    I_{i}(t) =
     & \frac{e}{h} \int d E \, d E^{\prime} \exp \left(\mathrm{i} \frac{E - E^{\prime}}{\hbar} t \right)                                    \\
     & \qquad \times \left[b_{i \mathrm{e}}^{\dagger}(E) b_{i \mathrm{e}}(E') - b_{i \mathrm{h}}^{\dagger}(E) b_{i \mathrm{h}}(E') \right],
  \end{aligned}
\end{equation}
where \(i=3,4\), indicating that the current is measured at \(\mathrm{D}_3\) or \(\mathrm{D}_4\), respectively.
The QPCb acting as a beam splitter is described by an energy-independent scattering matrix:
\begin{equation}\label[equation]{eq.Sb}
  \left(
  \begin{array}{c}
      b_{3\alpha} \\
      b_{4\alpha}
    \end{array}
  \right)
  = S_{\mathrm{b}}
  \left(
  \begin{array}{c}
      b_{1\alpha} \\
      b_{2\alpha}
    \end{array}
  \right), \quad
  S_{\mathrm{b}} =
  \begin{pmatrix}
    \cos \theta  & \sin \theta \\
    -\sin \theta & \cos \theta
  \end{pmatrix},
\end{equation}
where $\alpha= \mathrm{e}, \mathrm{h}$ and the indices 3, 4 represent the drains to which the wave packet is emitted.

Both two sources $\mathrm{S}_{1,2}$ are grounded, ensuring zero current measured at the drains $\mathrm{D}_{3,4}$, \emph{i.e.}
\(\langle \psi_{\mathrm{out}} | I_{3,4}(t) | \psi_{\mathrm{out}} \rangle=0\).
However, the current fluctuation is nonzero, and the associated shot noise between outgoing channels \(i\) and \(j\) is defined as the Fourier transform of the symmetrized current-current correlation function as
\[
  S_{ij}(\omega) = \int d t~\mathrm{e}^{\mathrm{i} \omega t }  \left\langle \psi_{\mathrm{out}} | \Delta I_i(t) \Delta I_j(0) + \Delta I_j(0) \Delta I_i(t) | \psi_{\mathrm{out}} \right\rangle,
\]
where \(\Delta I_j(t) = I_j(t) - \left\langle I_j(t) \right\rangle\). In the calculaiton of both current and shot noise, the fermionic anticommutation relations are utilized:
$
  \{b_{i}(E), b_{j}^{\dagger}(E') \} = \delta_{ij} \delta(E - E'), \,
  \{b_{i}(E), b_{j}(E') \} = 0,
  \{b_{i}^{\dagger}(E), b_{j}^{\dagger}(E') \} = 0.
$
We focus on the zero-frequency noise power at low temperatures (assumed to be zero). Zero bias voltage indicates that the Fermi sea makes no contribution to the noise poweer, which arises entirely from the electron-hole pairs above the vacuum $|\bar{0}\rangle$. A straightforward derivation
yields the noise power determined by the \( g \) matrix and the scattering matrix \( S_{\mathrm{b}} \) as
\[
  \mathcal{S}=S_{33} = S_{44} = - S_{34} = \frac{2 e^{2}}{\tau} \left( \left| G_{34} \right|^2 + \left| G_{43} \right|^2 \right),
\]
where \(\tau = 2\pi/\omega\) is the driving period and
\begin{equation}\label[equation]{eq.Gmatrix0}
  G_{ij} \equiv \sum_{m,n} \tilde{g}_{ij} (S_{\mathrm{b}})_{mi} (S^{\dagger}_{\mathrm{b}})_{jn}, \quad
  \tilde{g}_{ij} = g_{ij}(\alpha \rightarrow \tilde{\phi}).
\end{equation}
Inserting Eq.~(\ref{eq.Sb}) into the above expression yields
\begin{equation}
  \begin{aligned}
    \mathcal{S}=\frac{4 e^2}{\tau}
    \left|
    \rho_{\mathrm{E}}\left(
    \cos\tilde{\phi}+\mathrm{i}\sin\tilde{\phi}\cos2\theta
    \right)-\mathrm{i} \rho_{\mathrm{N}} \sin 2 \theta
    \right|^2
  \end{aligned}\label{Shotnoise1}
\end{equation}
The first term arises from the contribution of the entangled state $|E\rangle$ as indicated by its amplitudes $\rho_{\mathrm{E}}$. Its dependence on the phase $\tilde{\phi}$ clearly reveals how the entanglement in Eq.~\eqref{eq.stateEntangle} modulates the bunching/antibunching behaviors of the quasiparticles. The second term, with amplitude $\rho_{\mathrm{N}}$, reflects a nonzero contribution from the normal state $|N\rangle$ composed of locally excited electron-hole pairs.
If $\rho_{\mathrm{E}}=0$, this contribution is proportional to the product of the transmission probability $\sin^2 \theta$ and the reflection probability $\cos^2\theta$, typical for uncorrelated particles~\cite{blanter2000shot}.

Notably, for an ideal beam splitter with \(\theta = \pi/4\) that corresponds to equal transmission and reflection probabilities, the entangled and normal states contribute separately to the noise power and Eq.~\eqref{Shotnoise1} reduces to
\begin{equation}\label{eq.noise.result}
  \mathcal{S}=\mathcal{S}_{\mathrm{E}} + \frac{4 e^2}{\tau} \rho_{\mathrm{N}}^2,\ \ \  \mathcal{S}_{\mathrm{E}}=\frac{2 e^2}{\tau} \rho_{\mathrm{E}}^2 \left(1 + \cos 2 \tilde{\phi}\right).
\end{equation}
In this case, the second term due to the normal state is independent of \(\tilde{\phi}\) and contributes solely as a constant background.
The modulation to the bunching/antibunching effects due to the entangled Nambu spinor is manifested as the $\tilde{\phi}$ dependence of the fitst term $\mathcal{S}_{\mathrm{E}}$. Specifically, for the triplet state with \(\tilde{\phi} = 0\) that leads to antibunching of quasiparticles, the noise power reaches its maximum, in stark contrast to scenarios involving entanglement in spin or orbital degrees of freedom. Meanwhile, for the singlet state with \(\tilde{\phi} = \pi/2\), the bunching behavior results in $\mathcal{S}_{\mathrm{E}}=0$. These findings align with our previous analysis, where the exchange symmetries of the Nambu spinor govern the bunching and antibunching of quasiparticles, while the shot noise exhibits an opposite charge correlation effect.


To facilitate experimental observation of the entangled Nambu spinor, it is preferable to have $\rho_{\mathrm{E}}\gg\rho_{\mathrm{N}}$, ensuring that the amplitude of \(|E\rangle\) dominate over that for \(|N\rangle\) in \(|\Psi_{\mathrm{out}}\rangle\).
This condition can be realized by carefully selecting appropriate parameters for the QPCa. To this end, the unitary matrix \(s_0\) is parameterized as
\[
  s_0 = \begin{pmatrix}
    re^{i\theta_{r}}            & \mathrm{i} te^{-i\theta_{t}} \\
    \mathrm{i} te^{i\theta_{t}} & re^{-i\theta_{r}}
  \end{pmatrix},
\]
where \( R = r^2 \) and \( T = t^2 \) represent the reflection and transmission probabilities, respectively, with \( R + T = 1 \). All parameters \( r \), \( t \), \( \theta_{r} \) and \( \theta_{t} \) are functions of gate voltage \(V_0\).
Referring to Appendix~\ref{sec.appendix1}, \(\rho_{\mathrm{E}}\) and \(\rho_{\mathrm{N}}\) are given by
\[
  \begin{aligned}
     & \rho_{\mathrm{N}} = \frac{V_{1}}{2} \left(R \partial_{V_0} \theta_{r} - T \partial_{V_0} \theta_{t}\right), \\
     & \rho_{\mathrm{E}} = \frac{V_{1}}{2} \left(
    \frac{
      \left(
      R \partial_{V_{0}} T - T \partial_{V_{0}} R
      \right)^2
    }{4 RT}+ RT \left( \partial_{V_0} \theta_{r} + \partial_{V_0} \theta_{t} \right)^2\right)^{\frac{1}{2}},
  \end{aligned}
\]
which are proportional to the driving voltage $V_1$.
Their ratio can be expressed as
\[
  \frac{\rho_{\mathrm{E}}}{\rho_{\mathrm{N}}}
  = \sqrt{
    \frac{1 + 4 R^{2} T^{2} \left( \partial_{T} \theta_{r} + \partial_{T} \theta_{t} \right)^2}
    {4 R T \left( R \partial_{T} \theta_{r} - T \partial_{T} \theta_{t} \right)^2}
  }.
\]
When QPCa approaches the transparent limit (\(R \to 0\)) and, meanwhile, \(\partial_{T} \theta_{t}\) is finite, then \(\rho_{\mathrm{E}} \gg \rho_{\mathrm{N}}\). In this scenario, \(\mathcal{S} \simeq \mathcal{S}_{\mathrm{E}}\), giving rise to pure signature of entanglement.


The high tunability of the setup allows for the calibration of relevant parameters through conventional electron transport before measuring the entangled quasiparticles.
To achieve an ideal beam splitter at QPCb, first pinch off QPCa and apply a voltage bias to the source $\mathrm{S}_1$. The scattering probabilities of QPCb, tuned by the local gate, can then be inferred from the currents measured at $\mathrm{D}_3$ and $\mathrm{D}_4$, with equal currents being the desired outcome. A similar procedure can be used to measure the scattering coefficients at QPCa and their dependence on $V_0$ by pinching off QPCb.
Subsequently, by opening the tunneling at QPCa, the entire setup functions as an interferometer. By adjusting the gate voltage \(V_g\) and observing the periodic oscillation of the current at \(\mathrm{D}_3\) or \(\mathrm{D}_4\), the relation between the phase $\tilde{\phi}$ and $V_g$ can be extracted.
The last but most important experimental consideration is the synchronization of quasiparticle arrival times at QPCb. Although the path lengths in Fig.~\ref{fig.1}(c) may appear unequal, it is imperative that particle-hole pairs generated at QPCa arrive simultaneously at QPCb.
This synchronization requirement arises because Hanbury Brown--Twiss-type interference critically depends on the temporal overlap of quasiparticles at the beam splitter. Experimentally, this can be achieved by using lateral gates to reshape the edge states and fine-tune their path lengths, as demonstrated in Ref.~\cite{roulleauDirectMeasurementCoherence2008}.
On the other hand,
once the path lengths are adjusted to be equal, the additional phase in the Nambu spinor caused by path length differences will also be eliminated simultaneously.
Once these parameters are calibrated, the setup is ready to measure the shot noise caused by entangled Nambu spinors.

In conclusion, we propose an effective method for detecting entangled Nambu spinors. Due to the entanglement encoded in the electron-hole degree of freedom, the correspondence between the bunching/antibunching behavior of the quasiparticles and the current correlation is reversed compared to conventional entanglement in spin or orbital degrees of freedom.
Compared to the previous scheme~\cite{luoEntanglementNambuSpinors2022}, the current proposal does not require coupling the edge states to superconductors, thus significantly simplifying the experimental implementation. Our work uncovers an unconventional correlation effect induced by quasiparticle entanglement, which may lead to interesting applications in quantum information processing.
\begin{acknowledgments}
  This work was supported by
  the National Natural Science Foundation of
  China (No.  12222406), the Natural Science Foundation of Jiangsu Province (No. BK20233001),
  the Fundamental Research Funds for the Central Universities (No. 2024300415), and
  the National Key Projects for Research and Development of China (No. 2022YFA1204701).
\end{acknowledgments}

\appendix
\section{g matrix}\label[section]{sec.appendix1}
Here, we discuss the properties of the \( g \) matrix within the framework of Floquet scattering theory \cite{moskaletsDissipationNoiseAdiabatic2002b,moskaletsFloquetScatteringTheory2002a}. First, the driving external field at QPCa is given by
\begin{equation*}
  V(t) = V_0 + V_1 \left( e^{\imi \omega t} + e^{-\imi \omega t} \right)/2.
\end{equation*}
In the perturbative limit where \( V_1 \rightarrow 0 \), the Floquet scattering matrix can be approximated as:
\begin{equation*}
  s(t) \approx {s} + {s}_{-} e^{\imi \omega t} + {s}_{+} e^{-\imi \omega t}.
\end{equation*}
As outlined in the main text, within the framework of the adiabatic approximation, we have \( {s}_{\pm} \equiv {s}_1 = \frac{V_1}{2} \left( \frac{\partial {s}_0}{\partial V_0} \right) \).
This scattering matrix satisfies the unitarity condition:
\begin{equation*}
  s(t) s^{\dagger}(t) = s^{\dagger}(t) s(t) = \hat{1}.
\end{equation*}
This implies that, accurate up to the first-order term in \( V_1 \), \( {S}(t) \) must satisfy the condition:
\begin{equation*}
  \begin{aligned}
     & s^{\dagger}(t)s(t) \approx s^{\dagger}_{0}s_{0} + 2 \cos(\omega t)\left( s^{\dagger}_{0}s_{1} + s^{\dagger}_{1}s_{0} \right) = \hat{1}; \\
     & s(t)s(t)^{\dagger} \approx s_{0}s^{\dagger}_{0} + 2 \cos(\omega t)\left( s_{0}s^{\dagger}_{1} + s_{1}s^{\dagger}_{0} \right) = \hat{1}.
  \end{aligned}
\end{equation*}
Thus, on the one hand, \( s_0 \) satisfies the unitarity requirement \( s^{\dagger}_{0}s_{0} = s_{0}s^{\dagger}_{0} = 1 \);
on the other hand, the following condition holds:
\begin{equation*}
  s^{\dagger}_{0}s_{1} + s^{\dagger}_{1}s_{0} = 0.
\end{equation*}
According to the definition in Eq.~(\ref{eq.gMatrix}), the \( g \) matrix must satisfy the following:
\begin{equation*}
  g = - g^{\dagger}; \quad
  \begin{pmatrix}
    g_{11} & g_{12} \\
    g_{21} & g_{22}
  \end{pmatrix}
  = - \begin{pmatrix}
    g^{*}_{11} & g^{*}_{21} \\
    g^{*}_{12} & g^{*}_{22}
  \end{pmatrix}.
\end{equation*}

For the off-diagonal elements, if we denote \( g_{12} = \rho_{\mathrm{E}} \mathrm{e}^{\mathrm{i} \alpha} \), then \( g_{21} = -g_{12}^{*} = - \rho_{\mathrm{E}} \mathrm{e}^{- \mathrm{i} \alpha} \).
For the diagonal elements, their specific form is constrained by the condition \( {s}_{1} = V_1 \left( \frac{\partial {s}_0}{\partial V_0} \right) / 2 \).
Given that \( {s}_{0} \) is a \( 2 \times 2 \) unitary matrix, it can be represented as:
\begin{equation*}
  {s}_{0} \equiv
  \begin{pmatrix}
    r & -t^{*} \\
    t & r^{*}
  \end{pmatrix},
\end{equation*}
and the form of \( {s}_{1} \) can be expressed as:
\begin{equation*}
  {s}_{1} = \frac{V_1}{2} \frac{\partial {s}_0}{\partial V_0}
  = \frac{V_1}{2}
  \begin{pmatrix}
    \partial_{V_0} r & - \partial_{V_0} t^{*} \\
    \partial_{V_0} t & \partial_{V_0} r^{*}
  \end{pmatrix}
\end{equation*}
Therefore:
\begin{equation*}
  \begin{aligned}
    g
     & \equiv s^{\dagger}_{0} s_{1}
    = \frac{V_1}{2}
    \begin{pmatrix}
      r^{*} & t^{*} \\
      -t    & r
    \end{pmatrix}
    \begin{pmatrix}
      \partial_{V_0} r & - \partial_{V_0} t^{*} \\
      \partial_{V_0} t & \partial_{V_0} r^{*}
    \end{pmatrix} \\
     & = \frac{V_1}{2}
    \begin{pmatrix}
      r^{*} \partial_{V_0} r + t^{*} \partial_{V_0} t
       & - r^{*} \partial_{V_0} t^{*} + t^{*} \partial_{V_0} r^{*} \\
      - t \partial_{V_0} r + r \partial_{V_0} t
       & t \partial_{V_0} t^{*} + r \partial_{V_0} r^{*}
    \end{pmatrix}
  \end{aligned}
\end{equation*}
This gives an additional relationship between \( g_{11} \) and \( g_{22} \):
\begin{equation*}
  g_{11} = g_{22}^{*}
\end{equation*}
Considering the anti-Hermiticity of \( {g} \), we obtain:
\begin{equation*}
  g_{11} = g_{22}^{*} = - g_{22}
\end{equation*}
Thus, we find that \( g_{11} \) and \( g_{22} \) are purely imaginary and have opposite signs. Therefore, we can define:
\begin{equation*}
  g_{11} \equiv \rho_{\mathrm{N}} \imi, \quad
  g_{22} \equiv - \rho_{\mathrm{N}} \imi
\end{equation*}

In summary, the final result yields a \( g \) matrix in the form of Eq.~(\ref{eq.gMatrixFinal}).

\end{document}